\documentclass[aps,prb,twocolumn,superscriptaddress,floatfix]{revtex4}
\usepackage{amsmath,amssymb,graphicx,graphics,color}
\usepackage{dcolumn}
\usepackage{bm}

\def\Xint#1{\mathchoice
{\XXint\displaystyle\textstyle{#1}}%
{\XXint\textstyle\scriptstyle{#1}}%
{\XXint\scriptstyle\scriptscriptstyle{#1}}%
{\XXint\scriptscriptstyle\scriptscriptstyle{#1}}%
\!\int}
\def\XXint#1#2#3{{\setbox0=\hbox{$#1{#2#3}{\int}$ }
\vcenter{\hbox{$#2#3$ }}\kern-.6\wd0}}

\def\dashint{\mathop{\Xint-}}

\begin{document}

\title{Resonances in a dilute gas of magnons and metamagnetism of isotropic frustrated ferromagnetic spin chains}

\author{M. Arlego}

\affiliation{Departamento de F\'{\i}sica, Universidad Nacional de La Plata, C.C.\ 67, 1900 La Plata, Argentina}

\author{F. Heidrich-Meisner}

\affiliation{Physics Department and Arnold Sommerfeld Center for Theoretical Physics, Ludwig-Maximilians-Universit\"at M\"unchen, 80333 M\"unchen, Germany}

\author{A. Honecker}

\affiliation{Institut f\"ur Theoretische Physik, Georg-August-Universit\"at G\"ottingen,
  37077 G\"ottingen, Germany}

\author{G. Rossini}

\affiliation{Departamento de F\'{\i}sica, Universidad Nacional de La Plata, C.C.\ 67, 1900 La Plata, Argentina}

\author{T. Vekua}

\affiliation{Institut f\"ur Theoretische Physik, Leibniz Universit\"at Hannover, 30167~Hannover, Germany}

\date{May 27, 2011; revised November 22, 2011} 

\begin{abstract}
We show that spin-$S$ chains with SU(2)-symmetric, ferromagnetic 
nearest-neighbor and frustrating antiferromagnetic next-nearest-neighbor 
exchange interactions exhibit metamagnetic behavior under the influence of 
an external magnetic field for small $S$, in the form of a first-order 
transition to the fully polarized state. The corresponding magnetization 
jump increases gradually starting from an $S$-dependent critical value of 
exchange couplings and takes a maximum in the vicinity of a ferromagnetic Lifshitz 
point. The metamagnetism results from resonances in the dilute magnon gas 
caused by an interplay between quantum fluctuations and frustration.
\end{abstract}

\maketitle

\section{Introduction}
\label{sec:intro}

Quantum spin systems at low temperatures share many of the macroscopic 
quantum behaviors with systems of bosons such as Bose-Einstein 
condensation, superfluidity (see Ref.~\onlinecite{giamarchi08} and 
references therein), or macroscopic quantum tunneling, \cite{chudnovsky88} 
and may indeed be viewed as quantum simulators of interacting 
bosons.\cite{giamarchi08} Spin systems with frustration, in particular, 
realize exotic phases of strongly correlated bosons, such as spin liquids 
\cite{balents10} or supersolids.\cite{liu73} Experimentally, such systems 
can be studied both in low-dimensional quantum magnets (see, e.g., 
Ref.~\onlinecite{balents10}) and in ultra-cold atomic 
gases.\cite{struck11,jo11} In the latter case, interest in the many-body 
physics has been excited by the extraordinary control over interactions 
between bosons via Feshbach resonances. \cite{bloch08} These can, in 
particular, be used to tune interactions from repulsive to attractive. In 
the attractive regime, the collapse of an ultra-cold gas of bosons was 
observed in experiments.\cite{gerton00} In this work, we discuss a 
mechanism by which the same can be achieved in spin systems.

In the large-$S$ limit, spins map onto  bosons with a finite but large Hilbert space, justifying a description
in terms of soft-core bosons. Yet, in many cases, similarities between spins and soft-core bosons
may even exist for $S=1/2$. In three dimensions (3D),  systems of spins
and bosons resemble  each other the more the smaller  the density of bosons is, whereas 1D spin-$S$ antiferromagnets close to saturation behave as spinless fermions, or hard-core bosons.
In spin systems, the  external magnetic field $h$  tunes the density of magnons. The limit of a dilute gas of magnons
is then realized as the fully polarized state (the vacuum of magnons) is approached from below.

In this work, we  argue that resonances can play a crucial role in frustrated quantum spin systems largely determining their low-energy behavior in an external magnetic field.
We  consider the frustrated ferromagnetic (FM) spin-$S$ chain and show that upon changing system parameters such as
the coupling constants or $h$, one can tune  the effective interaction between magnons from repulsive to attractive by exploiting the existence of resonances.
As a main result, we demonstrate that, close to resonances, where the scattering length is much larger than the lattice constant, and in the
case of  attractive effective interactions, the intrinsic hard-coreness of spins does not play a significant role in the  limit of a dilute gas of magnons.
A  behavior resembling the collapse of attractively interacting bosons \cite{nozieres82,Mueller}
therefore exists in such spin systems close to their fully polarized state.
The thermodynamic instability of collapsed states causes jumps in the magnetization curve just below saturation.
This has to be contrasted with the scattering length being of the order of a few lattice sites. In that case, which is realized for spin 1/2, the  formation of
mutually repulsive, multi-magnon bound states can be observed (see, e.g., Refs.~\onlinecite{vekua07,kecke07,hikihara08,sudan09,hm09}).

Our work demonstrates that   the mapping of a {\it purely 1D} spin system close to saturation  to an
effective theory of a dilute Bose gas properly accounts for the physics of the model. This is accomplished by
connecting the scattering vertices of the microscopic lattice model
with the coupling constants of the effective theory. Furthermore, despite the purely 1D nature
of our problem, a $1/S$ expansion is  a valuable tool and yields the correct physics.


Concretely, we study the following system:
\begin{equation}
H_S = \sum_{i=1}^L \left \lbrack J\vec{S}_i\cdot \vec{S}_{i+1}+ J'   \vec{S}_i\cdot \vec{S}_{i+2}-h  S^z_i \right\rbrack\,.
\label{eq:ham}
\end{equation}
$\vec{S}_i=(S_i^x, S^y_i, S^z_i)$ is a spin-$S$ operator acting on site $i$
and $L$ is the number of sites.
$J < 0$ is the FM, nearest-neighbor exchange interaction and $J'=1$ is the antiferromagnetic (AFM),
next-nearest-neighbor exchange interaction setting the energy scale. In the
absence of an external field $h$,
$H_S$
 has a FM ground state for $J\le-4$
 for all $S$ (see Ref.~\onlinecite{bader79}).
 $J=-4$ is a ferromagnetic Lifshitz point where the
quadratic term in the dispersion of the magnons vanishes.

Systems with competing FM and AFM interactions are of  timely interest,  \cite{shannon06} in particular, the
spin-$1/2$ version of Eq.~\eqref{eq:ham},\cite{chubukov91a,kolezhuk05,hm06a,vekua07,kecke07,hikihara08,sudan09} motivated by  the  experimental realizations in, e.g.,
 LiCuVO$_4$ (Refs.~\onlinecite{enderle05,enderle10}) and Li$_2$ZrCuO$_4$ (Ref.~\onlinecite{drechsler07}).
 The 1D case of   $J<0$ and $S > 1/2$  is largely unexplored  (for $J>0$ and $S>1/2$, see Refs.~\onlinecite{kolezhuk05} and \onlinecite{hm07}).

We are mainly interested in the region $-4<J<0$ and magnetization $M=S^z/{(S L)}$ ($S^z=\sum_i \langle S^z_i \rangle $) close
to saturation $M=1$, for general spin $S$. We will proceed in three steps:
First, in Sec.~\ref{sec:two_magnon}, we discuss the solution of the two-magnon problem and introduce the scattering length.
Second, in Sec.~\ref{sec:dilute}, we  map the low-energy limit of  Eq.~\eqref{eq:ham}, close
to saturation, to a dilute 1D gas of two species of bosons interacting via an effective short-range interaction.\cite{Batyev84,nikuni95}
We then calculate the interaction vertices in this effective theory using a $1/S$ expansion.
Finally, in Sec.~\ref{sec:dmrg}, we compare the analytical
results with  exact numerical ones using the density matrix renormalization group (DMRG) method \cite{white92b,schollwoeck05} and exact diagonalization (ED).
We put a particular focus  on the case of $S=1$.
A summary of our results is presented in Section \ref{sec:summary}, while
technical details of the mapping to a dilute gas and of the $1/S$ expansion are given in
Appendix \ref{app:dilute}. A comparison of DMRG results for open boundary conditions vs.\
results for periodic boundary conditions is shown in Appendix \ref{app:dmrg}.


\section{Two-magnon problem}
\label{sec:two_magnon}

\subsection{Solution of the two-magnon problem}

We now solve  the interacting
two-magnon problem, starting with the thermodynamic limit (for  $S=1/2$,  see Ref.~\onlinecite{gochev74}). On a chain of finite length $L$ with periodic boundary
conditions the total momentum $K$ is a good quantum number due to the
translational invariance of the Hamiltonian $H_S$ in Eq.\ \eqref{eq:ham}.
 Thus, it is convenient to use a basis separating momentum subspaces
\begin{equation}
|K,r\rangle=\sum_{l=1}^{L}e^{iK(l+r/2)}S_{l}^{-}S_{l+r}^{-}|F\rangle \, ,
\label{eq:mixed-basis}
\end{equation}
where $|F\rangle$  is the fully polarized state, $K=2q\pi/L$
($q=0,1,\cdots,L-1$) and $r$ is the relative distance of two magnons.
The allowed values of $r$ depend on $S$ and the parity of $L$ and $q$. For instance, in the case of $S>1/2$ and $L$ even,  $r=0, 1,..,L/2-1,(L/2)$ for $q$ odd (even).

We expand a general two-magnon state with momentum $K$
into the (unnormalized) basis of Eq.~(\ref{eq:mixed-basis}) as
\begin{equation}
|\Psi_{2M}\rangle=\sum_{r}C_{r}|K,r\rangle
\end{equation}
and determine $C_{r}$ analytically by solving the two-magnon Schr\"odinger
equation
\begin{equation}
H_S|\Psi_{2M}\rangle=E_{2M}|\Psi_{2M}\rangle.
\end{equation}
This leads to the recurrence relations
\begin{eqnarray}
\label{eq:recurrence}
\Omega_0C_0&=&\frac{S}{\sqrt{S(2S-1)}}(\zeta_{1}C_1 +\zeta_{2}C_2  )\nonumber\\
(\Omega_0-J)C_1&=&\frac{(2S-1)^{3/2}}{{S^{3/2}}}\zeta_{1}C_0+\zeta_{1}C_2 \nonumber\\
&&+\zeta_{2}(C_1+C_3) \nonumber\\
(\Omega_0-1)C_2&=& \frac{(2S-1)^{3/2}}{{S^{3/2}}}\zeta_{2}C_0+\zeta_{2}C_4\nonumber\\
&&+ \zeta_{1} (C_1+C_3)\nonumber\\
\Omega_{0}C_{r} & = & \zeta_{1}\left(C_{r+1}+C_{r-1}\right)\nonumber\\
&&+  \zeta_{2}\left(C_{r+2}+C_{r-2}\right),\quad\mathrm{for}\,\,r\geq3,
\end{eqnarray}
where $\zeta_{1}=2SJ\cos{(K/2)}$, $\zeta_{2}=2S\cos{(K)}$.
When $|\Psi_{2M}\rangle$ is a bound state, $\Omega_0=E_b-4S(1+J^2/8)$ where $E_b$ is the (negative) binding energy
(defined as the bound-state energy minus the energy of the minimum of the two-magnon scattering states).

The (unnormalized) two-magnon  bound states for a given $K$ are constructed
with the ansatz
\begin{equation}
C_{r}=e^{-\kappa_{-}r}+ve^{-\kappa_{+}r}\quad(r\geq1),\label{eq:Ansatz}
\end{equation}
which, inserted in Eq.~(\ref{eq:recurrence}),  leads to a characteristic quartic equation  for $r\geq 3$
\begin{equation}
\Omega_{0}z^{2}-\zeta_{1}(z^{3}+z)-\zeta_{2}(z^{4}+1)=0 \, ,
\end{equation}
$z$ being  any of $e^{-\kappa_{\pm}}$ with $\text{Re}[\kappa_{\pm}]>0$.
The remaining unknown quantities $C_{0}$, $v$ and $E_b$ are determined from the remaining relations listed in Eq.~(\ref{eq:recurrence}).

\subsection{Scattering length in the lattice problem}

 For $S>1/2$, bound states with energies below the minimum of the two-magnon scattering continuum exist only for $K\simeq \pm 2k_{cl}$ and only in a finite window of couplings
\begin{equation}\label{range_J1c}
    -4 < J < J_{cr}(S),
\end{equation}
with $S$-dependent critical values $J_{cr}(S)$, as illustrated in Table~\ref{tab:Jcr}.
The critical value $J_{cr}(S)$, which is $J_{cr}\approx -2.11$ for $S=1$,  quickly approaches  $J_{cr}(S) \simeq -4$
with increasing $S$ (see Tab.~\ref{tab:Jcr}). In fact, the $1/S$ analysis to be presented in Sec.~\ref{sec:dilute} suggests the existence
of an $S_{cr}$ beyond which this window disappears completely.

\begin{table}[b]
\begin{tabular}{| c | c | c|c |c |}
\hline
  $S$ & 1  & 3/2  &2  & 5/2  \\\hline
 $-J_{cr}(S)$ &  2.11 (2.95) & 3.31 (3.42) &3.68 (3.66)   &3.84 (3.80) \\ \hline
\end{tabular}
\caption{Critical  exchange couplings $J_{cr}(S)$ for the existence of metamagnetism in  Eq.~\eqref{eq:ham}
derived from solving the two-magnon problem (values in parenthesis:  Results from the $1/S$ expansion of Sec.~\ref{sec:dilute}).}
\label{tab:Jcr}
\end{table}
We define the  \emph{scattering length} of bound states, in the thermodynamic limit $L \rightarrow \infty$,
from their spatial extent (in analogy to the continuum problem of particles interacting via a short-range, attractive potential):
\begin{equation}
\label{boundstatescatteringlength}
a_S =\frac{1}{\mathrm{min} \{\text{Re}[\kappa_{\pm}]\}}.
\end{equation}
The binding energy takes its lowest value ({\it i.e.}, the largest  absolute value) for $K=K^*\simeq \pm 2k_{cl}=\pm 2\arccos{(-J/4 )}$  and this quantity, with extremely high accuracy, is related to the scattering length by
\begin{equation}
\label{bindingenergy}
E_b(K^*)\simeq-\frac{1}{ma_S^2}\, ,
\end{equation}
where $m$ is the one-magnon \emph{mass},
\begin{equation}
\label{magnonmass}
m=\frac{2}{S(4-J)(4+J)}.
\end{equation}
The relation  Eq.~(\ref{bindingenergy}) between the binding energy and the
scattering length that holds for our microscopic lattice model is typical
for a 1D Bose gas in the continuum interacting via an attractive contact potential, the Lieb-Liniger model.\cite{LL}

\begin{figure}
\includegraphics[width=0.9\columnwidth]{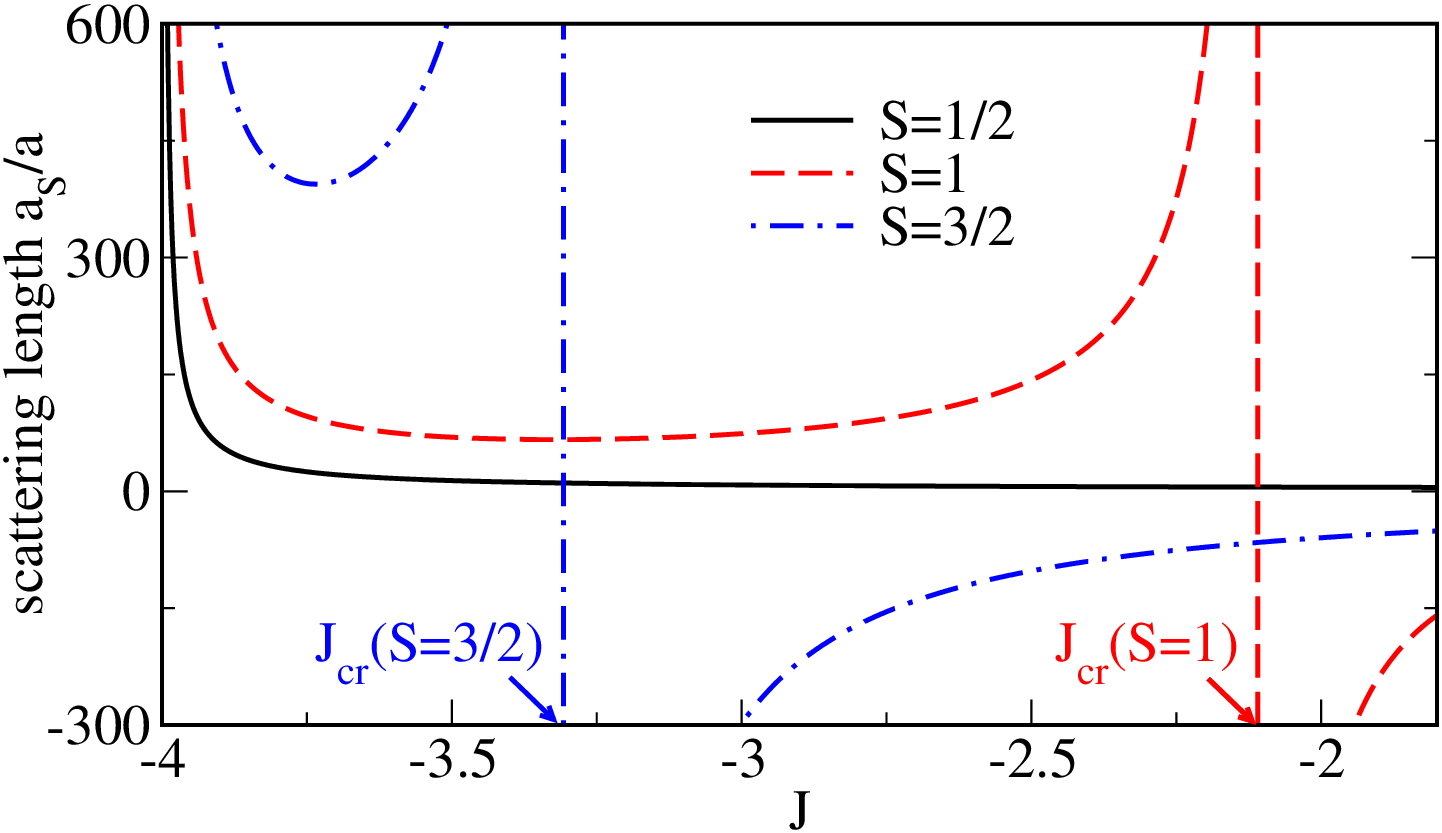}
\caption{(Color online) 1D scattering length $a_S/a$ ($a$: lattice spacing)  for $S=1/2,1,3/2$ (solid, dashed, dot-dashed line). }
\label{fig:scatt}
\end{figure}

The scattering length $a_S$ can also be determined from the scattering problem of two magnons for general $S$ in the thermodynamic limit.
We have  solved this problem, with the momenta of the two magnons (which participate in scattering) taken in the vicinity of the same dispersion minimum,
$k_1=k_{cl}+k$ and $k_2=k_{cl}-k$.
From the asymptotic form of the two magnon scattering state wavefunction we extract the scattering phase shift $\delta_S (k)$ for any $S$,
\begin{equation}
 \lim_{r\to \infty}C_r \sim \cos{(rk+\delta_S(k))}.
\end{equation}
To extract the scattering length from the scattering phase shift we use the same relation as in a 1D continuum model of particles interacting via
a short-range potential
\begin{equation}
\label{scatteringstatescatteringlength}
 a_S=\lim_{k\to 0} \frac{\cot{(\delta_S(k))}}{k}.
\end{equation}
This allows us to calculate the scattering length in the repulsive regime
$a_S < 0$ as well (when two-magnon bound states are not formed below the
minimum of the scattering continuum). In the attractive regime $a_S > 0$,
the scattering lengths obtained from both approaches [{\it i.e.},
Eq.~(\ref{boundstatescatteringlength}) and Eq.~(\ref{scatteringstatescatteringlength})] are in excellent agreement with each other.
As a side note on terminology, we call attractive (repulsive) regime the one in which the effective interaction between magnons
is attractive (repulsive).

We can now generalize the procedure\cite{Okunishi98} of mapping the antiferromagnetic (unfrustrated) spin-$S$
chain close to saturation onto the  low-density limit of
the Lieb-Liniger model with a coupling constant
\begin{equation}
\label{okunishi}
g_0=-\frac{2}{
ma_S} \, . 
\end{equation}

However, in our model the single-magnon dispersion has  two minima. The effective theory thus will be a two-component (two species) Lieb-Liniger model.
There are two types of low-energy scattering processes, first when momenta of two magnons
are in the vicinity of the same dispersion minimum that we have presented above
(intraspecies scattering), and second, when the momenta $k_1$ and $k_2$ of two magnons are in the
vicinity of different minima of dispersion, {\it i.e.}, $k_1=k_{cl}+k$ and $k_2=-k_{cl}-k$
(interspecies scattering). For the latter case we can repeat all steps presented
above and extract another coupling constant $\tilde g_0$ from the interspecies scattering length, $\tilde a_S$ in analogy with Eq.~(\ref{okunishi}).
We obtain that $\tilde g_0>0$ (implying that bound states with total momentum
$K=0$ are never formed below the scattering continuum), and $\tilde
g_0>g_0$ for any $S$ in the region $-4<J<0$. Since the relation $\tilde g_0>g_0$ always holds, the relevant scattering length at low energies is the intraspecies scattering length $a_S$.

The scattering length, shown in Fig.~\ref{fig:scatt}, can be  well described, for small $S>1/2$,
by a sum of two terms (resonances):
$a_S\simeq \lambda^-_S/(4+J)+\lambda^+_S/(J_{cr}(S)-J)$ [where $\lambda^{\pm}_S$ are numerical prefactors].
We emphasize that, for $S\geq 1$, the scattering length is in general  much larger than the lattice spacing, as is evident from Fig.~\ref{fig:scatt}:
For $S=1$, $a_S$ takes a minimum  at $J\simeq -3.3$ with $ a_S\simeq 80a$.
Additionally, the emergence of bound states manifests itself by a diverging scattering length at $J_{cr}(S)$,
where $a_S$ changes its sign jumping from $-\infty$ to $+ \infty$.
Thus, bound states are typically shallow, with a binding energy given by Eq.~(\ref{bindingenergy}). In addition, the minima in their dispersion occur at incommensurate momenta $K^*$.

For $S=1/2$, any $J<0$ induces a two-magnon bound state with total
momentum $K=\pi$, \cite{chubukov91a,hm06a} {\it i.e.}, $a_S > 0$ and there
is no resonance at $-4 < J<0$ for $S=1/2$. Hence, $S=1/2$ is very
different from the $S>1/2$ case where bound states with $K=\pi$ are never
below the two-magnon scattering continuum, and, as discussed above, a
resonance exists for $-4 < J<0$ and $1\leq S < S_{cr}$.

In order to analyze the finite-size effects with respect to results
in the thermodynamic limit, we have numerically diagonalized $H_S$ in
the basis given in Eq.~(\ref{eq:mixed-basis}) (see Ref.~\onlinecite{kecke07} for
details of the procedure). We  obtained the full spectrum,
\textit{i.e.}, the scattering continuum
and bound/antibound states if present, for selected values of $J \in[-4,0]$
in systems with up to $L=4000$ sites and several values of $S$.
Table \ref{tab:Num_Jcr}  shows the numerical determination of $J_{cr}(S)$ for $S=1$ and different system sizes.
Although finite-size effects are apparent, a quadratic fit in $1/L$ to numerical data for $J_{cr}(S)$ extrapolates to $J_{cr}\simeq -2.11$ in agreement
with the result determined directly in the  thermodynamic limit (see the preceding discussion and Table \ref{tab:Jcr}).

\begin{table}[tb!]
\begin{center}
\begin{tabular}{| c | c | c |c |c |}
\hline
  $L$ & 1000  & 2000  & 4000  & $\infty$  \\\hline
  $-J_{cr}(S=1)$ & $2.25$ & $2.16$ & $2.13$  & $2.11$ \\ \hline
\end{tabular}

\caption{Finite-size dependence of critical values $J_{cr}(S)$ for the emergence of bound states below the minimum
of the two-magnon continuum of scattering states, for $S=1$.}
\label{tab:Num_Jcr}
\end{center}
\end{table}


\section{Mapping of the spin Hamiltonian to  a dilute  gas of bosons}
\label{sec:dilute}

In this section, we describe our effective theory in the thermodynamic limit, for  the case of a finite
(though vanishingly small) density of magnons.
The mapping to a dilute gas of bosons is motivated by the following observation: For $S>1/2$, we have shown that $a_S$ is large. Hence in the dilute limit,
we can safely neglect the hard-core constraint and take the continuum limit.
For $S=1/2$, on the contrary, the scattering length $a_S$ is typically of the order of a few lattice constants and only for $-4<J<-3.9$ does $a_S$ become
comparable to the smallest value of the scattering length for $S=1$.

For $S\geq 1$, close to saturation, and in the low-energy limit, we therefore map our system onto a dilute two-component gas of bosons interacting with effective short-range interactions.
Many-body effects will be incorporated by properly shifting the two-body T matrix off-shell as explained in Ref.~\onlinecite{Lee2002}.
We show that, while the {\it inter}species interaction is always repulsive and
stronger than the {\it intra}species interaction, the latter undergoes a sign
change. When the intraspecies interaction becomes negative, the  bosons are unstable
against  a collapse.
We show that a $1/S$ expansion captures this physics
correctly and, similar to the case of the (unfrustrated) Heisenberg chain, \cite{Johnson}
is applicable to the present problem, albeit its one-dimensional nature.

\subsection{Effective Hamiltonian}
Using the Dyson-Maleev transformation \cite{DysonMaleev} (the Dyson-Maleev representation is used here for convenience. We have checked
that the explicitly hermitean Holstein-Primakoff representation, \cite{HolsteinPrimakoff} to leading order $1/S$,
provides equivalent results)
\begin{eqnarray}
S_i^z&=&S-a_i^{\dagger}a_i \, ,\quad S_i^+=  \sqrt{2S}a_i \, , \nonumber\\
S_i^-&=&\sqrt{2S}a_i^{\dagger}(1-a_i^{\dagger}a_i/2S)\,,
\end{eqnarray}
we map  Eq.~\eqref{eq:ham} onto a bosonic problem:
\begin{equation}
\label{DysonMaleev}
H=\!\sum_k(2S\epsilon_k-\mu)a_k^{\dagger}a_k
+\!\!\!\sum_{k,k',q} \!\!\! \frac{\Gamma_0(q;k,k')}{2L} a_{k+q}^{\dagger}a_{k'-q}^{\dagger}a_{k}a_{k'} \, ,
\end{equation}
where
$$
\epsilon_{k}=J\cos{k}+\cos{2k}-(J\cos{k_{cl}}+\cos{2k_{cl}})\ge 0
$$ is the single-magnon dispersion and $k_{cl}=\arccos{(-J/4)}$.
Note that in our normalization, the minima of the single-particle dispersion are at zero energy: $\epsilon_{\pm k_{cl}}=0$.
The bare interaction vertex $\Gamma_0$  is given by
$\Gamma_0(q;k,k') = V_q-\frac{1}{2}(V_k+V_{k'})$ with
$V_k= 2J\cos{k}+2\cos{2k}$. The  chemical potential is $\mu=h^{cl}_s-h$,
where $h_s^{cl}$ is the classical saturation field value $h^{cl}_s=S(J+4)^2/4$.
We are interested in the dilute regime $\mu\to 0$.

Concentrating on the low-energy behavior we arrive at,
via a Bogoliubov procedure, \cite{Batyev84,nikuni95} a two-component Bose gas interacting via a $\delta$-potential
with Hamiltonian density
\begin{equation}
\mathcal{H}_{\rm eff}=\sum_{\alpha}-\frac{|\nabla \psi_{\alpha}|^2}{2m}+\frac{g_0(S)}{2}(n_1^2+n_2^2)+\tilde g_0(S) \, n_1 n_2\, .
\label{eq:effective}
\end{equation}
Here $\psi_\alpha$, $\alpha=1,2$ describe bosonic modes with momenta close to $\pm k_{cl}$
({\it i.e.}, the Fourier transforms of $\psi_\alpha$ are
$\psi_1(k\!\!\to\!\!0) \approx a_ {k_{cl}+k}$,
$\psi_2(k\!\!\to\!\!0) \approx a_ {-k_{cl}+k}$) and $n_{\alpha}=\psi_\alpha^{\dagger}\psi_\alpha$ are the corresponding densities.
The bare coupling constants of the effective 1D model of the two-component Bose gas, $g_0(S)$ and $\tilde g_0(S)$, are,  in the dilute limit of bosons,  related to the renormalized vertices of the microscopic model Eq.~(\ref{DysonMaleev}) through
\begin{eqnarray}
\label{microeffective}
&&  \Gamma(0;k_{cl},k_{cl})=  \frac{g_0(S)}{1+g_0(S){\sqrt{2m}}/({\pi \sqrt{\mu}})} ,\\
&&\Gamma(0;k_{cl},\!-k_{cl})+\Gamma(-2k_{cl};k_{cl},\!-k_{cl})\!=\!\frac{\tilde g_0(S)} {1+\frac{\tilde g_0(S){\sqrt{2m}}}{\pi \sqrt{\mu}}}. \nonumber
 \end{eqnarray}
The relations Eq.~(\ref{microeffective}) follow from a generalization of the corresponding equation for the case of a
one-component Bose gas \cite{kolomeisky92,Lee2002} to the two-component case using an RG analysis\cite{KolezhukRG}
(see Appendix \ref{app:dilute} for details of the calculation).

\subsection{$1/S$ expansion}
\label{subsect:1overS}

Next, we  apply a $1/S$ expansion to calculate the interaction vertices and extract the coupling constants $g_0(S)$ and $\tilde g_0(S)$.
Using a standard ladder approximation the Bethe-Salpeter equation for the vertices $\Gamma$ reads:
\begin{eqnarray}
\label{BetheSalpeterEquation}
\Gamma(q;k,k')&=& \Gamma_0(q;k,k')  \\
&-&\frac{1}{2SL}\sum_{p} \frac{ \Gamma_0(q-p;k+p,k'-p) } {\epsilon_{k+p}+\epsilon_{k'-p}} \Gamma(p;k,k')\nonumber\,.
\end{eqnarray}
Setting the transferred momentum $q=0$ in $\Gamma$ and the incoming momenta to $k=k'=k_{cl}$,
we get (see Appendix \ref{app:dilute} for details):
\begin{eqnarray}
\label{nine}
&&\Gamma(0;k_{cl},k_{cl})\left[1+\frac{V_0-V_{k_{cl}}}{2SL}  \sum_{p} \frac{ 1 } {\epsilon_{k_{cl}+p}+\epsilon_{k_{cl}-p}}\right]=\nonumber\\
&&V_0-V_{k_{cl}}+\frac{1}{2SL}\sum_{p}\! \left[1-\frac{  V_{p}-V_0 } {\epsilon_{k_{cl}+p}+\epsilon_{k_{cl}-p}} \right]\Gamma(p;k_{cl},k_{cl})\nonumber\\
&&-\frac{1}{2SL}\sum_{p} \frac{ (V_0-V_{k_{cl}}) \left[\Gamma(p;k_{cl},k_{cl})-\Gamma(0;k_{cl},k_{cl})\right] } {\epsilon_{k_{cl}+p}+\epsilon_{k_{cl}-p}}.\nonumber\\
\end{eqnarray}
Now, in the spirit of the $1/S$ expansion, we replace the renormalized vertex with the bare vertex on the right hand side of Eq.~(\ref{nine}),
$\Gamma(p;k_{cl},k_{cl})\to 2\epsilon_p$, which is possible  since there are no infrared divergences.
Regularizing the left hand side of Eq.~(\ref{nine}) as in Refs.~\onlinecite{kolomeisky92,Lee2002}, and using  Eq.~(\ref{microeffective})
we extract the coupling constants of the effective model.
The  analytical expression for $g_0(S)$ is
\begin{equation}
\label{LLC_wc1}
g_0(S)=\frac{F}{1-\frac{F(J^2-8)}{|J|S(16-J^2)^{3/2}}}
\end{equation}
(the derivation of Eq.~\eqref{LLC_wc1} can be found in Appendix \ref{app:dilute} and
the constant $F$ is given in Eq.~\eqref{eq:valueF}). To leading order in $J+4$,
\begin{equation}
\lim _{J\to -4^+} g_0(S) \simeq \frac{S-S_{cr}}{4S}(J+4)^2+O\left((J+4)^{5/2}\right)\,.
\end{equation}
To first order in $1/S$, we obtain $S_{cr}=6$, which is not that large a number, hence corrections beyond $1/S$ may affect $S_{cr}$.
In the same way we calculate $\tilde g_0(S)$ as
\begin{equation}
\label{g0tilde}
\tilde g_0(S)=\frac{\tilde F}{1+\frac{J^2-8}{16S}}
\end{equation}
(the derivation is presented in detail in Appendix~\ref{app:dilute}; see Eq.~\eqref{tildeF} for the expression for $\tilde F$).

From Eqs.~\eqref{LLC_wc1} and \eqref{g0tilde}, we notice that $g_0(S) < \tilde g_0(S)$ for $-4<J<0$.
Thus the state below saturation is a single-component one.
Provided the interactions are repulsive, the ground state is a translationally invariant chiral state,\cite{kolezhuk05}
where bosons prefer to `condense' at the same  minimum of the single-particle dispersion
since they experience a minimal repulsion  there.\cite{jackeli04}
We also note that $|g_0|m\ll 1$ for $J<0$, hence interactions between bosons are generically weak. 
In particular, even though $m\to \infty$ when approaching the ferromagnetic Lifshitz point, $|g_0|m\to 0$.

\begin{figure}
\includegraphics[width=0.9\columnwidth]{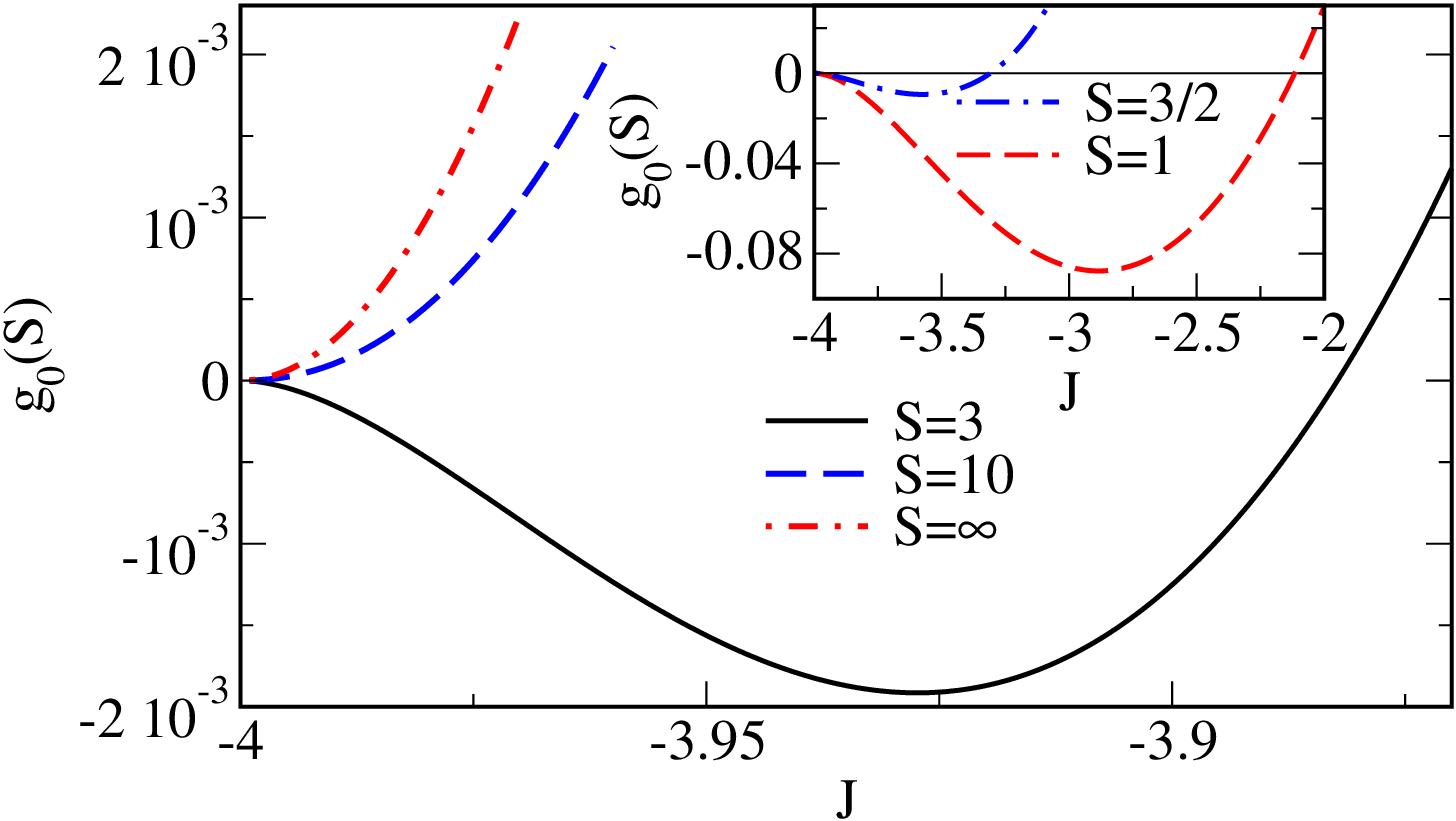}
\caption{(Color online) Effective bare intraspecies interaction $g_0(S)$, for $S=3$ (solid line), representative of the  generic behavior for $S<S_{cr}$,
$S=10$ (dashed curve), representative of $S>S_{cr}$ [dot-dashed line: $g_0(S=\infty)$]. Inset: $g_0$ for $S=1$ and $3/2$.
}
\label{fig:geff}
\end{figure}

The effective bare intraspecies interaction is depicted in Fig.~\ref{fig:geff} and behaves as
$g_0 (S) \sim [J-J_{cr}(S)]$ for $J\to J_{cr}(S)$. 
The scattering length is related to the effective coupling constant by Eq.~\eqref{okunishi}, signaling a resonance at $J_{cr}(S)$.
Thus, we see that for $S<S_{cr}$ there is a finite region near $J\simeq-4$ where $g_0 (S)<0$ and bosons attract each other, producing a collapsed state.

To corroborate this, using ED  for Eq.~\eqref{eq:ham} and $S=1$ with periodic boundary conditions,
we have calculated the ground-state momentum of the states with a small, but
finite number of magnons, which is incommensurate,  supporting the picture of a uniform chiral state in the repulsive case  $J>J_{cr}$,
and a collapsed state in the attractive case  $J<J_{cr}$ at one of the two minima of the
single-particle dispersion.

In the attractive case, $\partial^2E_0/\partial n^2<0$, where $E_0$ and $n $ are the bosons' ground-state energy and density, respectively.
In the language of spins, the inverse magnetic susceptibility at saturation becomes negative,
and hence, following standard arguments,\cite{dmitriev01} we conclude
that there is a first-order transition at $M=1$, {\it i.e.}, a
jump in the magnetization curve just below saturation.

As pointed out above, the case of $S=1/2$ is special since the scattering
length is typically of the order of the lattice constant here. The mapping of
the $S=1/2$ case to a two-component Bose gas (by the procedure presented
above for $S>1/2$) can be trusted only for $J \to -4$, where the
scattering length becomes much larger than the lattice constant. In that
case we can easily incorporate the exact hard-core constraint into our
formalism \cite{nikuni95} and again expect that $S=1/2$ also shows
metamagnetic behavior. This conclusion is in agreement with DMRG results
for $S=1/2$.\cite{sudan09} Note that a metamagnetic jump can also
be stabilized for spin $1/2$ with suitable anisotropic exchange interactions.\cite{GMK98,hirata99}
However, with our procedure we cannot account
for the formation of stable two-, three-, and four-magnon bound states that is
characteristic for most of the region $J>-4$ in the spin-$1/2$
frustrated ferromagnetic Heisenberg chain.\cite{hm06a,kecke07,hikihara08,sudan09}

Going back to $S>1/2$, at lower $M$, corresponding to higher densities of magnons, the hard-core nature of spins  eventually prevails
as well, resulting in a uniform ground state at a nonzero momentum.
However,
as already mentioned, from the finite-size analysis of the two-magnon problem, we observe that bound states disappear
with decreasing $L$, suggesting that the attractive effective potential (in the limit of a small magnon density) can become
repulsive upon increasing the magnon densities.  Thus, the state below the jump
({\it i.e.}, $0<M<1-\Delta M_{\mathrm{jump}}$,
where $\Delta M_{\mathrm{jump}}$ is the height of the jump)
will be similar to the one encountered in the case of $J>J_{cr}$, {\it i.e.}, it is a translationally uniform chiral state.

\begin{figure}[t!]
\includegraphics[width=0.85\columnwidth]{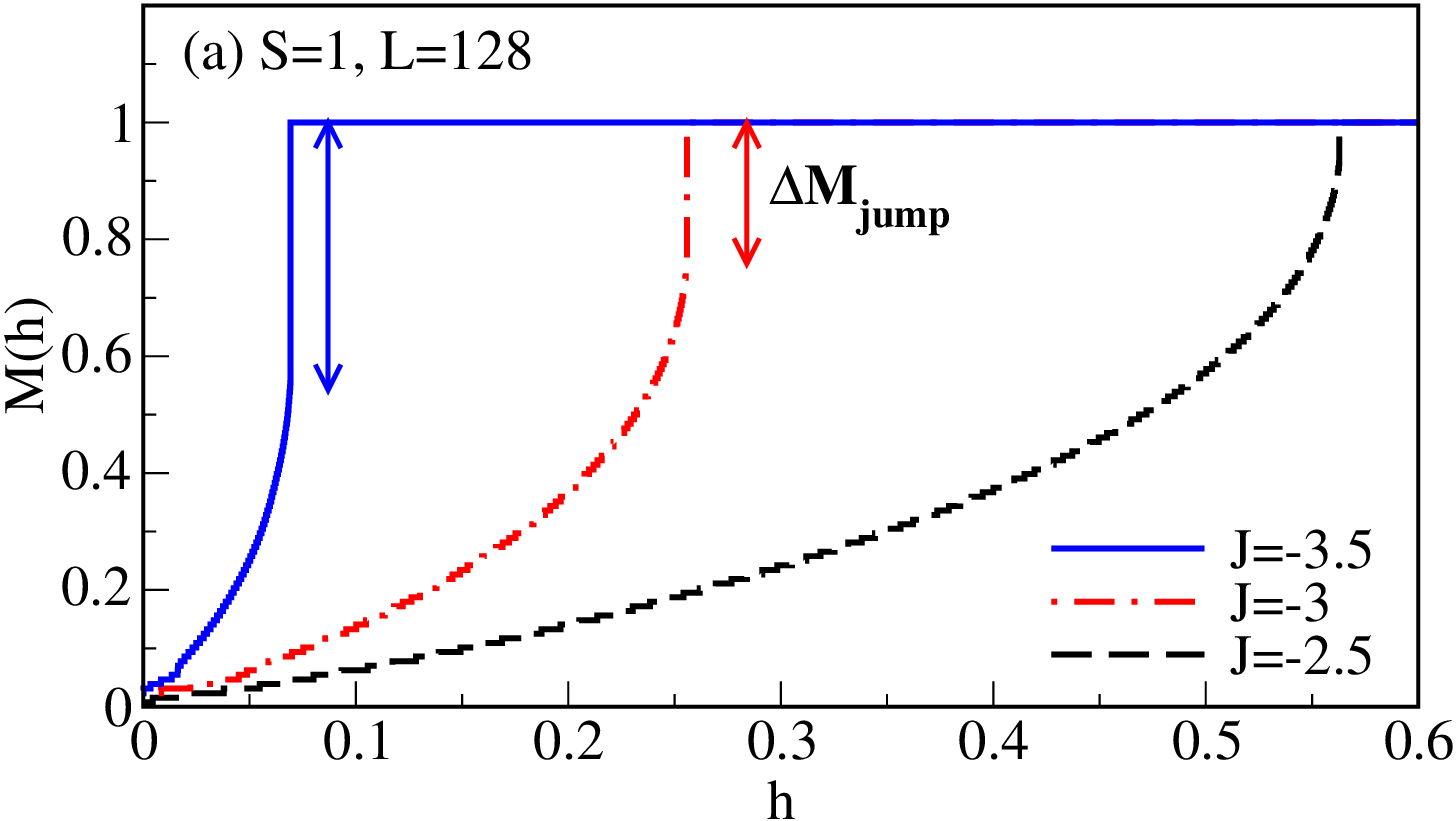} \\[1mm]
\includegraphics[width=0.9\columnwidth]{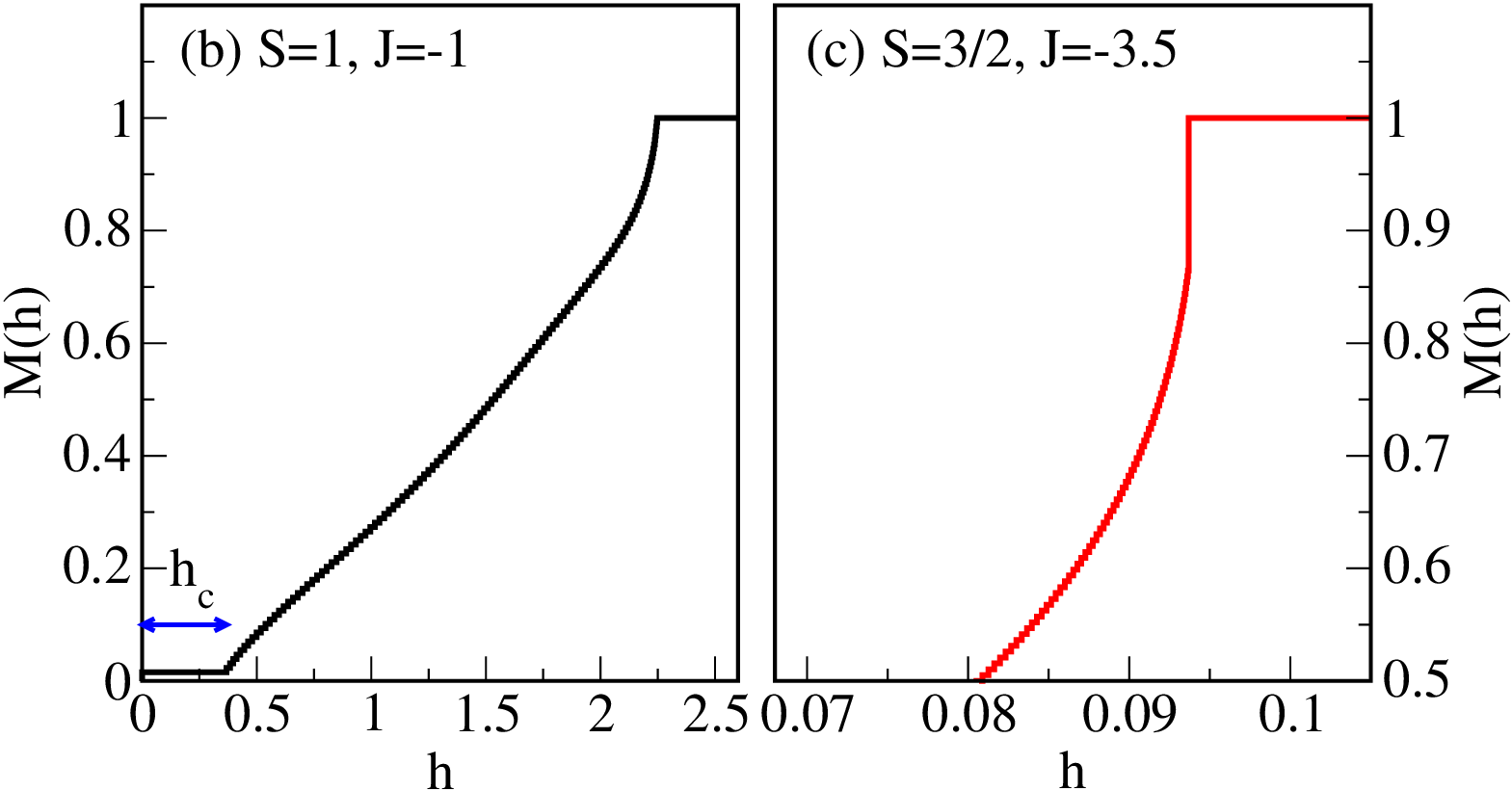}
\caption{(Color online) (a) Magnetization curves $M(h)$ for $S=1$ at $J=-2.5,-3,-3.5$.
(b) $M(h)$ for $S=1$, $J=-1$ 
(c) $M(h)$ for $S=3/2$ at $J=-3.5$
(all for $L=128$).}\label{fig:mag_h}
\end{figure}


\section{DMRG results}
\label{sec:dmrg}

Next, we turn to  numerical results for the case of  $S=1$ (unless stated otherwise), solving for the ground state of Eq.~\eqref{eq:ham}
in a finite magnetic field $h$, using ED where possible or DMRG.\cite{fn-alps}
We present data from DMRG simulations using up to $1200$  states,  for $L\leq 128$ sites, and for
open boundary conditions (OBC), unless stated otherwise.

\subsection{Magnetization curves and magnetization profiles}

\begin{figure}[t!]
\includegraphics[width=0.9\columnwidth]{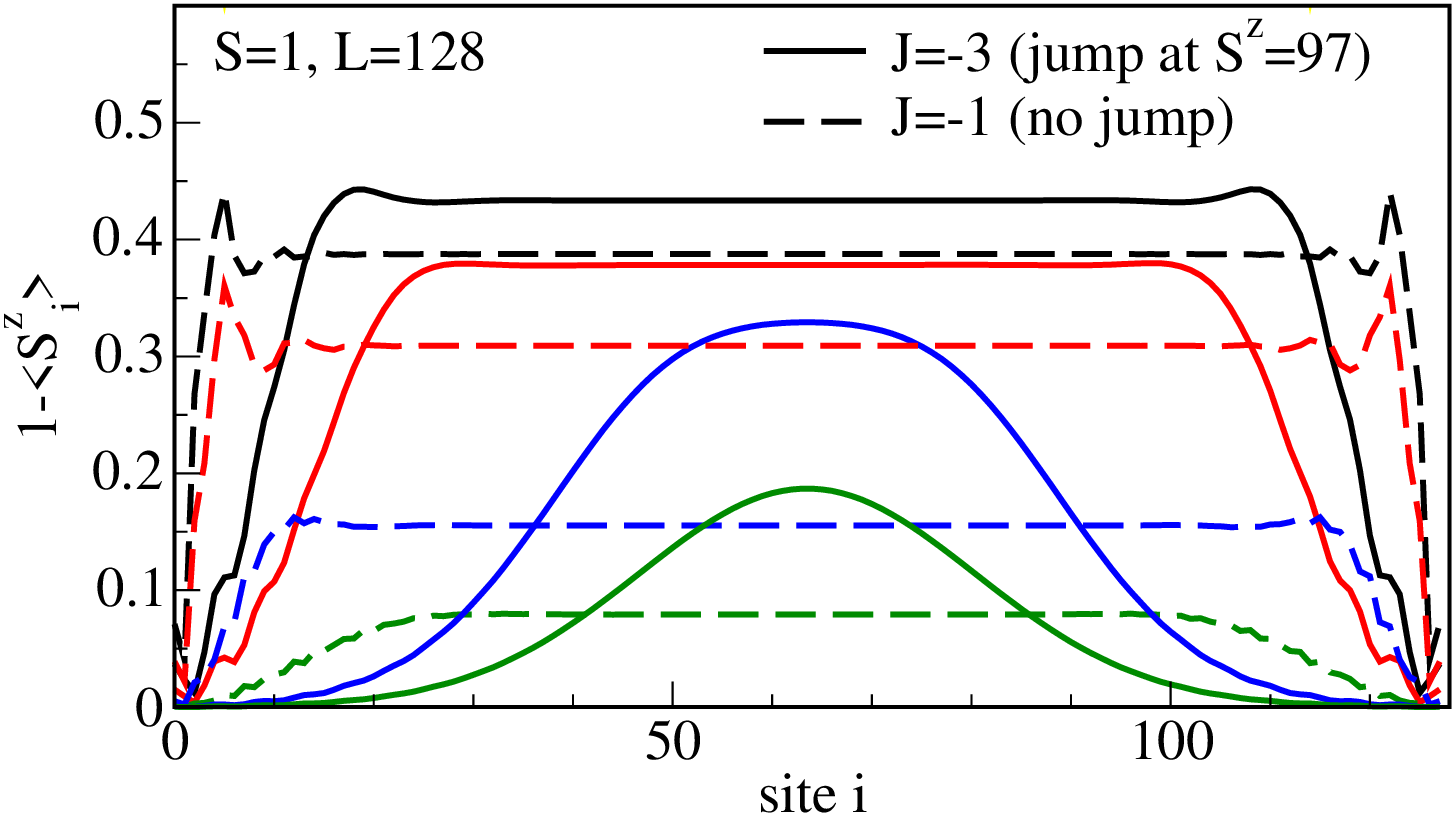}
\caption{(Color online) Magnetization profiles for $S=1$, $L=128$ at
$J=-1$ (no jump, dashed lines) and $J=-3$ (jump at $S^z=97$)
for $S^z=80,90,110,120$ (top to bottom).}\label{fig:mag_prof}
\end{figure}

The main result of this work, namely, the metamagnetic transition from a gapless finite-field phase to full saturation, is clearly
seen in the magnetization curves shown in Fig.~\ref{fig:mag_h}(a). For $S=1$, we observe the appearance of this jump for $-4< J \lesssim -2$,
while for $S=3/2$ [an example is shown in  Fig.~\ref{fig:mag_h}(c)] the jump exists in
a much narrower window $J\lesssim -3 $. This is consistent with our analytical results for $J_{cr}$, listed in Tab.~\ref{tab:Jcr}.
Moreover, for $-2 \lesssim J<0 $, we resolve a plateau in $M(h)$ at $M=0$, which is due to the Haldane gap\cite{whitehuse}
in this $S=1$ system
[see Fig.~\ref{fig:mag_h}(b)]. This gap defines the critical field $h_c$ that separates the gapped Haldane phase from the
finite-$M$ phase with a smooth $M(h)$-behavior for $h_c<h<h_{\mathrm{sat}}$.

The collapse of magnons manifests itself in the
magnetization profiles ($1-\langle S^z_i \rangle$ vs.\ site $i$) using OBC, displayed in Fig.~\ref{fig:mag_prof}   for  $J=-3$ and  $J=-1$. In the former
case, there is a jump, but in the latter, there is none.
Clearly, in the states that get skipped over ($S^z>97$ for $J=-3$),  magnons collapse into the center of the system, whereas for actual ground states below the metamagnetic transition, the magnetization profiles become flat.
By contrast, in the case of $J=-1$ where the transition to the fully polarized state is smooth and continuous, {\it all} profiles are, apart from boundary effects, flat.

\subsection{Central charge}

Our effective theory developed in Section \ref{sec:dilute}  suggests that the gapless phase 
in the region $h_{c}<h<h_{\mathrm{sat}}$ is a one-component phase (where $h_{\mathrm{sat}}$ is the saturation field).
To substantiate this result, one can make use of entanglement measures such as
the von-Neumann
entropy to extract the central charge, which directly yields the number of components of the gapless state.

The von-Neumann entropy is defined as
\begin{equation}
S_{vN}(l)=- \mbox{tr}(\rho_l \ln \rho_{l}) \, ,
\end{equation}
where $\rho_l$ is the reduced density matrix of a subsystem of length $l$ of our one-dimensional
chain of length $L$. In a gapless state that is conformally invariant, the $l$ and $L$ dependence of the von-Neumann entropy
is given by\cite{vidal03,calabrese}
\begin{equation}
S_{vN}(l) =\frac{c}{3} \ln\left( \frac{L}{\pi} \sin(\frac{\pi}{L} l)\right)+g\,,
\label{eq:cch}
\end{equation}
which is valid for systems with periodic boundary conditions (PBC). PBC are preferable for the calculation of the central charge from 
Eq.~\eqref{eq:cch} since for OBC, there may be additional oscillatory terms. $g$ is a non-universal constant that depends on $M$. As
DMRG directly accesses the eigenvalues of these reduced density matrices,\cite{schollwoeck05}
it is straightforward to measure $S_{vN}(l)$ with this numerical method.

\begin{figure}[tb!]
\includegraphics[width=0.9\columnwidth]{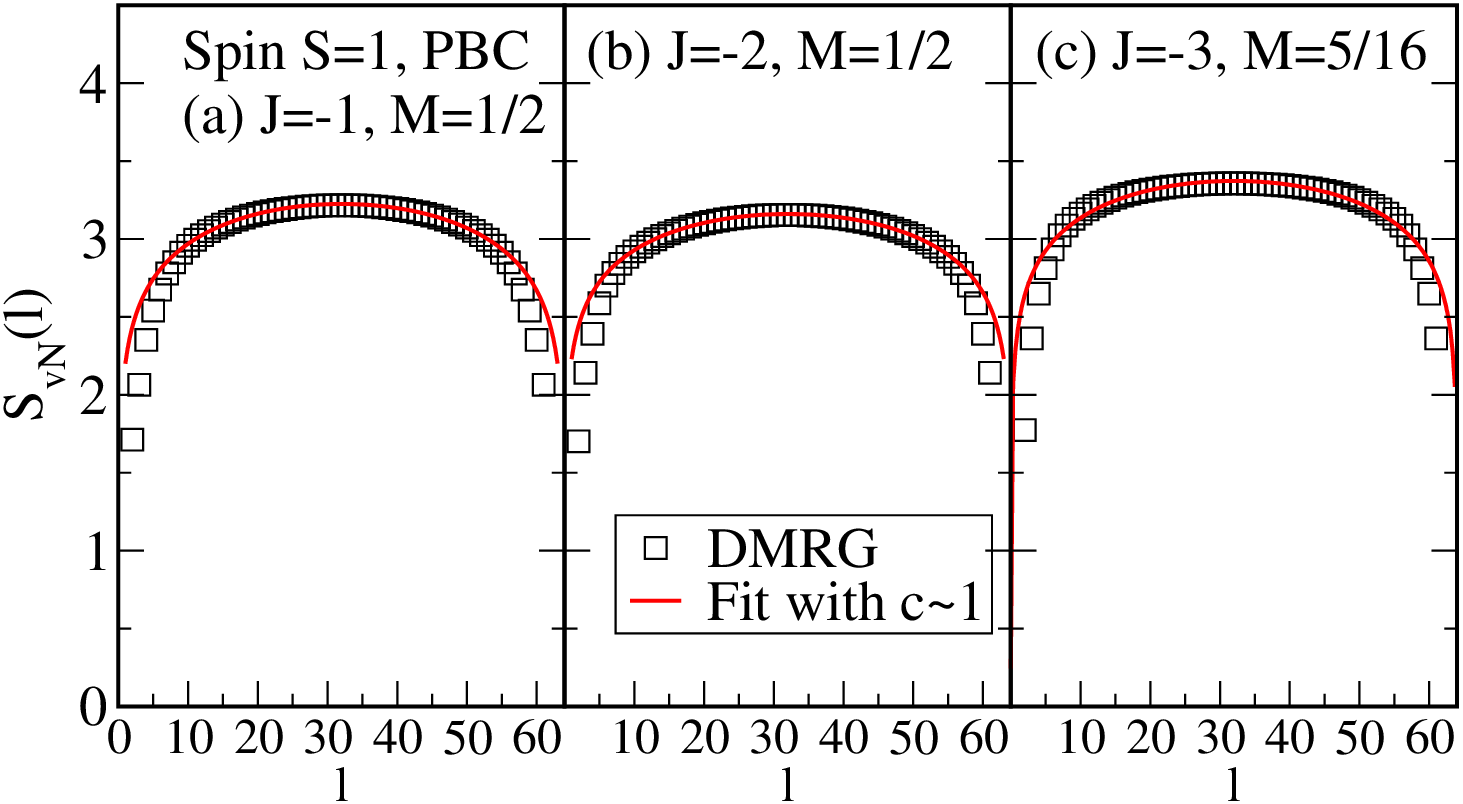}
\caption{(Color online)
DMRG results for the von-Neumann entropy $S_{vN}(l)$ in the gapless
phase $h_{c}<h<h_{\mathrm{sat}}$ of the $S=1$ system:
(a) $J=-1$, $M=1/2$, (b) $J=-2$, $M=1/2$, (c) $J=-3$, $M=5/16$ (symbols).
The lines are fits
to Eq.~\eqref{eq:cch}, resulting in $c=1.0\pm0.1$  in all cases (we exclude $S_{vN}(l)$ for $l< 10$ and $l>54$ from the fit). In this
figure, we display results for periodic boundary conditions and $L=64$ sites.
}\label{fig:svn}
\end{figure}

Some typical DMRG results (squares) for systems of $L=64$ and periodic boundary conditions are presented in Fig.~\ref{fig:svn}.
We have fitted the expression Eq.~\eqref{eq:cch} to our numerical data (shown as solid lines
in the figure) and obtain $c=1.0\pm 0.1$ in all examples. Therefore, we expect the
gapless phase to be a (chiral) one-component liquid. Note, though, that at both small $M$ and $|J|$,
where the convergence of DMRG is notoriously difficult, we cannot completely rule out the presence
of a $c=2$ region, which, however, is irrelevant for the main conclusions of our work.

\subsection{Phase diagram for $S=1$}

\begin{figure}
\includegraphics[width=0.9\columnwidth]{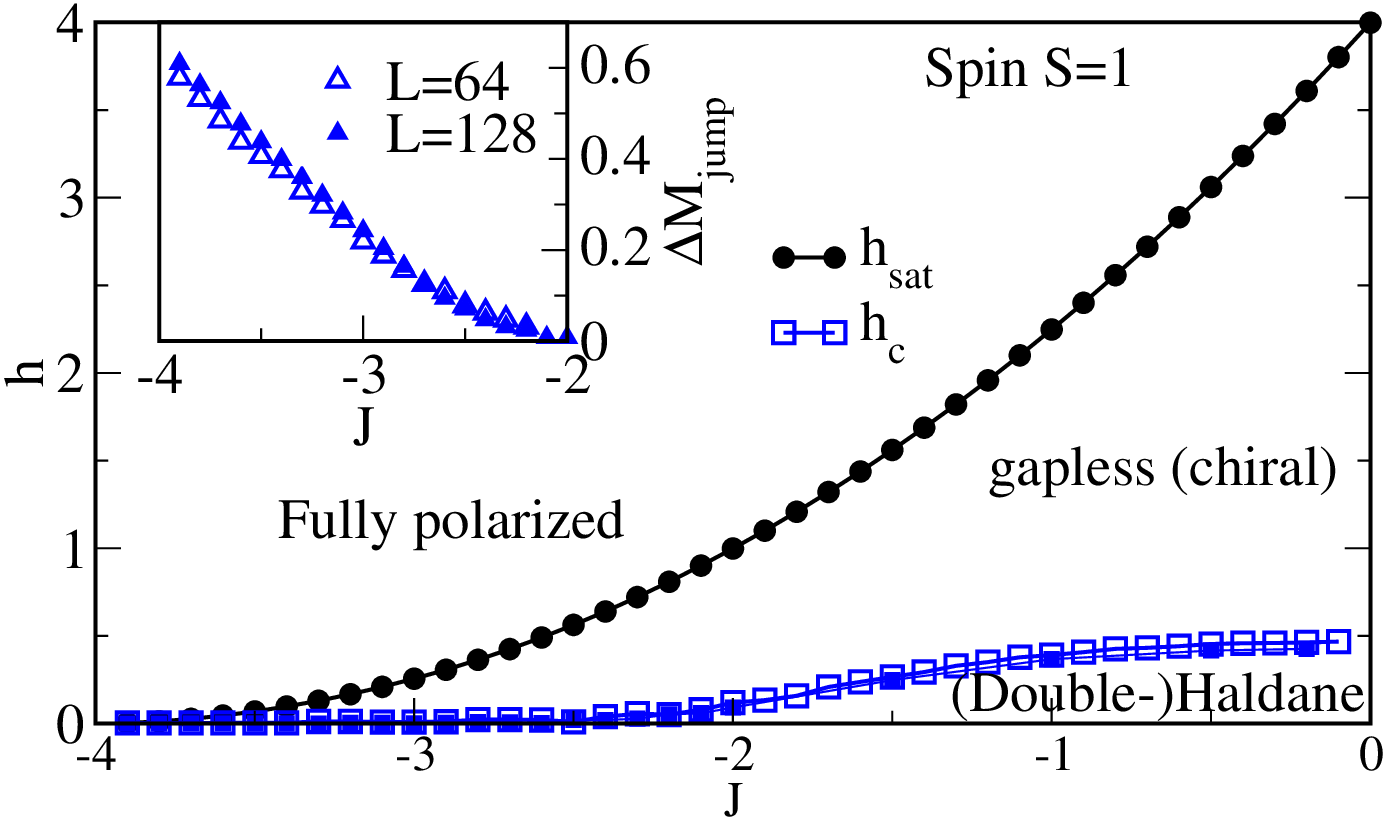}[t!]
\caption{(Color online) Phase diagram of Eq.~\eqref{eq:ham} for $S=1$  (circles: saturation field $h_{\mathrm{sat}}$; squares: spin gap $h_{c}$). Inset:
height $\Delta M_{\mathrm{jump}}$ of the metamagnetic jump vs.\ $J$. The comparison of $L=64$ (open symbols) and  $L=128$ (solid symbols) as well
as finite-size scaling (not shown here) supports  that both $\Delta M_{\mathrm{jump}}$ and $h_c$ are finite in  extended
regions of $J$.}\label{fig:phase}
\end{figure}

Our results for the $S=1$ chain are summarized in the $h$ vs.\ $J$ phase diagram Fig.~\ref{fig:phase}.
We identify three phases: (i) a gapped $M=0$ phase at $h<h_c$  (similar to the Double-Haldane phase known for $J>0$, see Ref.~\onlinecite{kolezhuk96}), (ii) a gapless (chiral) finite-field phase for $h_c<h<h_{\mathrm{sat}}$,
and (iii) the fully polarized state at $h_{\mathrm{sat}}<h$ (with $h_{\mathrm{sat}}=0$ for $J<-4$).

$\Delta M_{\mathrm{jump}}$  is plotted in the inset of Fig.~\ref{fig:phase}: the jump sets in at $J\lesssim -2 $
(close to where the  zero-field gap becomes small rendering it difficult to resolve it numerically), consistent with our theory.

Since in the limit of $J\to 0$, one has two spin-1 chains with antiferromagnetic
interactions which both separately have a Haldane gap at zero field,\cite{whitehuse} upon coupling the chains, one obtains the so-called Double-Haldane phase (in contrast to the regular Haldane phase that
is inherited from a single spin-1 chain with antiferromagnetic interactions). Both phases, Double-Haldane and Haldane phase, are
realized in the frustrated, antiferromagnetic spin-1 chain,\cite{kolezhuk96} yet in our case, only the Double-Haldane phase  exists.
The determination of the corresponding spin gap $h_c$ is a bit subtle.
Namely, a Haldane chain with open boundaries gives rise to spin-1/2
excitations at the open ends.\cite{kennedy90} Since we have two chains
(for small $|J|$), we have a total of four spin-1/2 end spins. Hence, the spin gap
in Fig.~\ref{fig:phase} is determined from
\begin{equation}
h_c = E(S^z=3) - E(S^z=2) \, ,
\label{eq:HcOBC}
\end{equation}
where $E(S^z)$ is the ground-state energy in a sector with a given total $S^z$.

It is worth emphasizing several differences with the phase diagram of the spin-1/2 version of Eq.~\eqref{eq:ham}.
First, for $S=1$, there are no multipolar phases, which occupy a large portion of the corresponding spin-1/2
phase diagram.\cite{hm06a,vekua07,kecke07,hikihara08,sudan09}
Second, the spin-1/2 system features an instability towards nematic order,\cite{chubukov91a,vekua07,kecke07} which can be
excluded on general grounds for integer spin-$S$ chains,\cite{kolezhuk05} even for  $0<M\ll 1$.

\section{Summary}
\label{sec:summary}

In conclusion, we showed that resonances can play a crucial role in 
determining the low-energy behavior of frustrated quantum spin systems 
subject to a magnetic field. The proximity of resonances caused by an 
interplay between frustration ($J>-4$) and quantum fluctuations ($1/2<S 
<S_{cr}$) results in extremely large values of the 1D scattering length 
that allows to develop an effective theory of a weakly interacting 
two-component Bose gas. The quasi-collapse of the dilute gas of magnons 
provides the physical origin of the emergent metamagnetism.

The preditictions of our analytic theory were verified by numerical data. We focused on the 
case $S=1$ since there the jump in the magnetization curve is the most 
pronounced. As a by-product, we obtained the phase diagram for the $S=1$ 
$J$-$J'$ chain with antiferromagnetic $J'$ and ferromagnetic $J$ in a 
magnetic field. This phase diagram is remarkably simple. In particular, there
are no multipolar phases in the case $S=1$, in marked contrast to the
spin-1/2 case.\cite{hm06a,vekua07,kecke07,hikihara08,sudan09}

\acknowledgments
We thank H. Frahm, A.\ Kolezhuk, R.\ Noack, and D. Petrov for fruitful discussions.
A.H.\ acknowledges  financial support from the DFG via
a Heisenberg fellowship (HO~2325/4-2).
T.V.\ is supported by the Center of Excellence QUEST.
G.R.\ and M.A.\ are partially supported by CONICET (PIP 1691) and ANPCyT (PICT 1426).

\appendix

\section{Details on the mapping to a dilute Bose gas}
\label{app:dilute}

In Sec.~\ref{sec:dilute}, we mapped the spin Hamiltonian Eq.~\eqref{eq:ham} close to saturation to
 an effective, bosonic field theory Eq.~\eqref{eq:effective}. Here we provide the details of generalizing the procedure of obtaining the coupling constants from the many-body T-matrix for an effective single-component Bose gas model, which is described in Ref.~\onlinecite{Lee2002},
 to the case relevant to us here, namely an effective theory of a two-component Bose gas.

We introduce the `mean-field interaction coefficients',
\begin{eqnarray}
\label{effectiveinteraction}
g(S)&=&\Gamma(0;k_{cl},k_{cl}) \\
\tilde g(S)&=&\Gamma(0;k_{cl},-k_{cl})+\Gamma(-2k_{cl};k_{cl},-k_{cl}) \, ,
\end{eqnarray}
where $\Gamma(q;k,k')$ is the full interaction vertex of the Hamiltonian Eq.~(\ref{DysonMaleev}).
They satisfy mean-field like relations for $\tilde g_{0}>g_0$,
\begin{equation}
 \mu=g(S)n
\end{equation}
and for $\tilde g_{0}<g_0$
\begin{equation}
 \mu=(g(S)+\tilde g(S))n/2,
\end{equation}
where $n=\langle\sum_{\alpha}n_{\alpha} \rangle\to0$ is the total density of bosons.
 To get the connection between $g(S)$ and $\tilde g(S)$ on the one hand and the bare coupling constants of the effective 1D model of a two-component Bose gas,
$g_0(S)$ and $\tilde g_0(S)$ on the other hand, we generalize the corresponding equation for the case of a one-component Bose gas \cite{Lee2002} to the two-component case, consistent with the $SU(2)$ symmetry of the RG fixed point of a dilute, two-component  Bose gas,\cite{KolezhukRG}
$\lim_{\mu\to 0}g(S)= \lim_{\mu\to 0} \tilde g(S)=\pi\sqrt{\mu}/\sqrt{2m}$,
\begin{eqnarray}
\label{microeffectiveapp}
&&  g(S)= \frac{g_0(S)}{1+g_0(S){\sqrt{2m}}/({\pi \sqrt{\mu}})},\\
&&\tilde g(S)\!=\! \frac{\tilde g_0(S)}{1+\tilde g_0(S){\sqrt{2m}}/({\pi \sqrt{\mu}})}. \label{microeffective_a}
 \end{eqnarray}
To zeroth order in $1/S$, we have
\begin{equation}
g(S=\infty)=V_0-V_{k_{cl}}=\frac{(J+4)^2}{4}
\end{equation}
and
\begin{equation}
 \tilde g(S=\infty)= V_{2k_{cl}}+V_0-2V_{k_{cl}}>g(S=\infty) \, .
\end{equation}
Thus, classically ($S\to \infty, m\to 0$), we have $\tilde g_{0}>g_0$ for $-4<J<0$.
We will show below that incorporating quantum fluctuations does not modify this relation.

In the following we will calculate the effective intraspecies and interspecies interaction constants,  to see 
how $1/S$ corrections modify $g_0(S)$ and $\tilde g_0(S)$.
%
In order to obtain $g(S)$ we have to set in $\Gamma (q,k,k')$ given by Eq.~(\ref{BetheSalpeterEquation}), $q=0$ and $k=k'=k_{cl}$,
\begin{eqnarray}
g(S)&=& V_0-V_{k_{cl}} +\frac{1}{2S L}\sum_{p} \Gamma_p\nonumber\\
&&-\frac{1}{2S L}\sum_{p} \frac{  V_{p}-V_0+V_0-V_{k_{cl}} } {\epsilon_{k_{cl}+p}+\epsilon_{k_{cl}-p}} \Gamma_p,
\end{eqnarray}
where we have denoted $ \Gamma(p;k_{cl},k_{cl})=\Gamma_p$.
After straightforward manipulations we obtain,
\begin{eqnarray}
\label{nineapp}
&&g(S)\left[1+\frac{V_0-V_{k_{cl}}}{2S L}  \sum_{p} \frac{ 1 } {\epsilon_{k_{cl}+p}+\epsilon_{k_{cl}-p}}\right]=\nonumber\\
&&\quad V_0-V_{k_{cl}} +\frac{1}{2S L}\sum_{p} \Gamma_p \left[1-\frac{  V_{p}-V_0
} {\epsilon_{k_{cl}+p}+\epsilon_{k_{cl}-p}}\right] \qquad \nonumber\\
&&\quad -\frac{1}{2S L}\sum_{p} \frac{ (V_0-V_{k_{cl}}) (\Gamma_p-\Gamma_0) }
{\epsilon_{k_{cl}+p}+\epsilon_{k_{cl}-p}} \, .
\end{eqnarray}
Now, on the right hand side of Eq.~(\ref{nineapp}), we plug in the zeroth order vertex (in the $1/S$ expansion),
 $\Gamma(p;k_{cl},k_{cl})\to 2\epsilon_p$ (which is possible because there are no infrared divergences anymore), and use that $V_p-V_0=2(\epsilon_p-\epsilon_0)$.
Equation~(\ref{nineapp}), after passing to infinite system size, becomes
\begin{eqnarray}
\label{unreg}
g(S)\left[1+\frac{F}{4\pi S} \int\limits_{-\pi}^{\pi} \frac{ 1 }
{\epsilon_{k_{cl}+p}+\epsilon_{k_{cl}-p}}\right]=F \, ,
\end{eqnarray}
where
\begin{eqnarray}
F&=& V_0- (1+\frac{1}{2S})V_{k_{cl}} - \frac{1}{\pi S}\int\limits_{-\pi}^{\pi}{\mathrm d}p \frac{   \epsilon_p (\epsilon_p-\epsilon_0) } {\epsilon_{k_{cl}+p}+\epsilon_{k_{cl}-p}} \nonumber\\
&&- \frac{ V_{0}-V_{k_{cl}} }{2\pi S}\int\limits_{-\pi}^{\pi}{\mathrm d}p \frac{  \epsilon_p-\epsilon_0 } {\epsilon_{k_{cl}+p}+\epsilon_{k_{cl}-p}}
\label{eq:valueF}
\end{eqnarray}
and we have used that
$V_0-V_{k_{cl}}=F+O(1/S)$.
According to the scheme that we follow the two-body T matrix must be calculated
off-shell, thus the denominator in Eq.~(\ref{unreg}) must be understood as
$\epsilon_{p}\to \epsilon_{p}+C\mu/4S$, where the exact value of the numerical constant
is $C=\pi^2/8$.\cite{Lee2002} This leads to (compare Eqs.~\eqref{microeffective}
and \eqref{effectiveinteraction})
\begin{equation}
\label{renormalizedpotential2}
g(S)= \frac{g_0(S)}{1+g_0(S){\sqrt{2m}}/({\pi \sqrt{\mu}})},
\end{equation}
where we have introduced the intraspecies Lieb-Liniger coupling constant $g_0(S)$ as in Eq.~(\ref{LLC_wc1}).
Note that all integrals presented in this section are evaluated analytically, which is a  nice feature of the $1/S$ treatment of our problem.

%

Now we outline the calculation of $\tilde g(S)$. We denote $\Gamma(p;k_{cl},-k_{cl})=\tilde \Gamma_p$ and obtain the Bethe-Salpeter equation for
\begin{equation}
\tilde \Gamma_0
=V_0 -V_{k_{cl}}   + \frac{1}{4\pi S}  \int\limits_{-\pi}^{\pi} {\mathrm d} p \tilde \Gamma_p
- \frac{1}{4\pi S}  \int\limits_{-\pi}^{\pi} {\mathrm d} p \frac{  V_{-p}-V_{k_{cl}} } {2\epsilon_{k_{cl}+p}}\tilde \Gamma_p
\nonumber
\end{equation}
and
\begin{eqnarray}
\tilde \Gamma_{-2k_{cl}}
&=&V_{-2k_{cl}} -V_{k_{cl}}   + \frac{1}{4\pi S}  \int\limits_{-\pi}^{\pi} {\mathrm d} p  \tilde \Gamma_p\nonumber\\
&&- \frac{1}{4\pi S}  \int\limits_{-\pi}^{\pi} {\mathrm d} p \frac{  V_{-2k_{cl}-p}-V_{k_{cl}} } {2\epsilon_{k_{cl}+p}}\tilde \Gamma_p .
\end{eqnarray}
Adding these two equations gives $\tilde g(S)=\tilde \Gamma_0+\tilde \Gamma_{-2k_{cl}}$,
\begin{eqnarray}
\label{gamma2}
\tilde g(S)
&=&V_0+V_{-2k_{cl}} -2V_{k_{cl}}   + \frac{2}{4\pi S}  \int\limits_{-\pi}^{\pi}
{\mathrm d} p  \tilde \Gamma_p \qquad \nonumber\\
&&- \frac{1}{4\pi S}  \int\limits_{-\pi}^{\pi} {\mathrm d} p \frac{
V_p+V_{-2k_{cl}-p} -2V_{k_{cl}} } {2\epsilon_{k_{cl}+p}}\tilde \Gamma_p.
\end{eqnarray}

We divide the last term in Eq.~(\ref{gamma2}) into two pieces,
\begin{equation}
-\frac{1}{4\pi S}  \int\limits_{-\pi}^{\pi} {\mathrm d} p \frac{
V_p+V_{-2k_{cl}-p}-2V_{k_{cl}} } {2\epsilon_{k_{cl}+p}}\tilde \Gamma_p=I_1+I_2
\, ,
\end{equation}
where
\begin{eqnarray}
I_1&=& - \frac{1}{4\pi S}  \int\limits_{-\pi-k_{cl}}^{-k_{cl}} {\mathrm d} p \frac{  V_p+V_{-2k_{cl}-p}-2V_{k_{cl}} } {2\epsilon_{k_{cl}+p}}\nonumber\\
&& \times (\tilde \Gamma_p- \tilde \Gamma_{-2k_{cl}}+\tilde \Gamma_{-2k_{cl}})
       \end{eqnarray}
and
\begin{eqnarray}
       I_2&=& - \frac{1}{4\pi S}  \int\limits_{-k_{cl}}^{\pi-k_{cl}} {\mathrm d} p \frac{  V_p+V_{-2k_{cl}-p}-2V_{k_{cl}} } {2\epsilon_{k_{cl}+p}}\nonumber\\
&& \times ( \tilde \Gamma_p- \tilde \Gamma_{0}+\tilde \Gamma_{0}) \, .
       \end{eqnarray}
Note that, for  convenience, we have shifted the first Brillouin zone,
$(-\pi,\pi)\to (-\pi-k_{cl},\pi-k_{cl})$. Shifting the T-matrix off-shell, we
have (the integral with a dash denotes the principal value)
\begin{eqnarray}
I_1&=& - \frac{1}{4\pi S}  \! \dashint\limits_{-\pi-k_{cl}}^{-k_{cl}}
  \!\!\!\! {\mathrm d} p \frac{  V_p+V_{-2k_{cl}-p}-2V_{k_{cl}} } {2\epsilon_{k_{cl}+p}}( \tilde \Gamma_p- \tilde \Gamma_{-2k_{cl}})\nonumber\\
&&- \frac{ \tilde \Gamma_{-2k_{cl}}}{4\pi S}  \int\limits_{-\pi-k_{cl}}^{-k_{cl}} {\mathrm d} p \frac{  V_p+V_{-2k_{cl}-p}-2V_{k_{cl}} } {2\epsilon_{k_{cl}+p}+C\mu/2S}
       \end{eqnarray}
and
\begin{eqnarray}
       I_2&=& - \frac{1}{4\pi S}  \int\limits_{-k_{cl}}^{\pi-k_{cl}} {\mathrm d} p \frac{  V_p+V_{-2k_{cl}-p}-2V_{k_{cl}} } {2\epsilon_{k_{cl}+p}}( \tilde \Gamma_p- \tilde \Gamma_{0})\nonumber\\
&&- \frac{ \tilde \Gamma_{0}}{4\pi S}  \int\limits_{-k_{cl}}^{\pi-k_{cl}} {\mathrm d} p \frac{  V_p+V_{-2k_{cl}-p}-2V_{k_{cl}} } {2\epsilon_{k_{cl}+p}+C\mu/2S}.
       \end{eqnarray}
The last terms in $I_1$ can be written as
\begin{eqnarray}
&-& \frac{ \tilde \Gamma_{-2k_{cl}}}{4\pi S}  \int\limits_{-\pi-k_{cl}}^{-k_{cl}} {\mathrm d} p \frac{  V_p+V_{-2k_{cl}-p}-2V_{k_{cl}} } {2\epsilon_{k_{cl}+p}+C\mu/2S}=\nonumber\\
&&- \frac{ \tilde \Gamma_{-2k_{cl}}}{4\pi S}
\dashint\limits_{-\pi-k_{cl}}^{-k_{cl}} {\mathrm d} p \frac{
V_p+V_{-2k_{cl}-p}-(V_0+V_{-2k_{cl}}) } {2\epsilon_{k_{cl}+p}}\quad\nonumber\\
&&- \frac{ \tilde \Gamma_{-2k_{cl}}}{4\pi S}  \int\limits_{-\pi-k_{cl}}^{-k_{cl}} {\mathrm d} p \frac{  V_0+V_{-2k_{cl}}-2V_{k_{cl}}} {2\epsilon_{k_{cl}+p}+C\mu/2S}.
       \end{eqnarray}
Similarly, for the last terms in $I_2$ we get,
\begin{eqnarray}
&-& \frac{ \tilde \Gamma_{0}}{4\pi S}  \int\limits_{-k_{cl}}^{\pi-k_{cl}} {\mathrm d} p \frac{  V_p+V_{-2k_{cl}-p}-2V_{k_{cl}} } {2\epsilon_{k_{cl}+p}+C\mu/2S}=\nonumber\\
&&- \frac{ \tilde \Gamma_{0}}{4\pi S}  \dashint\limits_{-k_{cl}}^{\pi-k_{cl}}
{\mathrm d} p \frac{  V_p+V_{-2k_{cl}-p}-(V_0+V_{-2k_{cl}}) }
{2\epsilon_{k_{cl}+p}}\quad\nonumber\\
&&- \frac{ \tilde \Gamma_{0}}{4\pi S}  \int\limits_{-k_{cl}}^{\pi-k_{cl}} {\mathrm d} p
\frac{  V_0+V_{-2k_{cl}}-2V_{k_{cl}}} {2\epsilon_{k_{cl}+p}+C\mu/2S} \, .
       \end{eqnarray}
Noting that
\begin{eqnarray}
&-& \frac{ 1}{4\pi S}  \dashint\limits_{-k_{cl}}^{\pi-k_{cl}} {\mathrm d} p \frac{  V_p+V_{-2k_{cl}-p}-(V_0+V_{-2k_{cl}}) } {2\epsilon_{k_{cl}+p}}\nonumber\\
&&=- \frac{ 1}{4\pi S}  \dashint\limits_{-\pi-k_{cl}}^{-k_{cl}} {\mathrm d} p
\frac{  V_p+V_{-2k_{cl}-p}-(V_0+V_{-2k_{cl}}) } {2\epsilon_{k_{cl}+p}} \quad \nonumber\\
&&=\frac{J^2-8}{16 S},
       \end{eqnarray}
and gathering all contributions, Eq.~(\ref{gamma2}) takes the form
\begin{eqnarray}
\label{gamma21}
&&\!\!\!\!\!\!\!\!\tilde g(S)\! \left[1\!+\!\frac{J^2-8}{16S}\! +\! \frac{V_0\!+\!V_{-2k_{cl}}-2V_{k_{cl}} }{16\pi S}  \!\int\limits_{-\pi}^{\pi} \!\!\frac{ {\mathrm d} p } {\epsilon_{p}+C\mu/4S}   \right]
\nonumber\\
&&\!\!\!\!=V_0+V_{2k_{cl}}-2V_{k_{cl}}+\frac{8+J^2}{4 S} \nonumber\\
&&\!\!\!\!-\frac{1}{4\pi S}\dashint\limits_{-\pi}^{0}{\mathrm d}p\frac{V_{p-k_{cl}}+V_{p+k_{cl}}-2V_{k_{cl}}}{2\epsilon_{p}}(\tilde \Gamma_{p-k_{cl}}-\tilde\Gamma_{-2k_{cl}})\nonumber\\
&&\!\!\!\!-\frac{1}{4\pi S}\! \int\limits_{0}^{\pi}\!\! {\mathrm d}p\frac{V_{p-k_{cl}}+V_{p+k_{cl}}-2V_{k_{cl}}}{2\epsilon_{p}}(\tilde \Gamma_{p-k_{cl}}-\tilde\Gamma_{0}).
\end{eqnarray}

Now we can plug the zeroth-order vertices into the right hand side of
Eq.~(\ref{gamma21}), in the spirit of the $1/S$ expansion: $\tilde\Gamma_{p}\to
2\epsilon_p$. Noting that $V_0+V_{-2k_{cl}} -2V_{k_{cl}} = \tilde F+O(1/S)$,
where
\begin{eqnarray}
\label{tildeF}
\tilde F&=& V_0+V_{2k_{cl}}-2V_{k_{cl}}+\frac{8+J^2}{4 S} \\
&-&\frac{1}{2\pi S}\dashint\limits_{-\pi}^{0}{\mathrm d}p\frac{V_{p-k_{cl}}+V_{p+k_{cl}}-2V_{k_{cl}}}{2\epsilon_{p}}(\epsilon_{p-k_{cl}}-\epsilon_{-2k_{cl}})\nonumber\\
&-&\frac{1}{2\pi S}\int\limits_{0}^{\pi}{\mathrm d}p\frac{V_{p-k_{cl}}+V_{p+k_{cl}}-2V_{k_{cl}}}{2\epsilon_{p}}(\epsilon_{p-k_{cl}}-\epsilon_{0}),\nonumber
\end{eqnarray}
in the same way as for $g(S)$ we obtain
\begin{equation}
\label{renormalizedpotential_2}
\tilde g(S)= \frac{\tilde g_0(S)}{1+\tilde g_0(S){\sqrt{2m}}/({\pi \sqrt{\mu}})}
\end{equation}
with $\tilde g_0(S)$ given in Eq.~\eqref{g0tilde}.

We see that in order $1/S$, quantum fluctuations do not modify the relation $\tilde g_0(S)>g_0(S)$ for $J<0$. Therefore, the interspecies interaction
is always positive (repulsive) for any $J$ and $S>1/2$, and behaves as $\tilde g_0(S)\sim (J+4)^2$ for $J\to -4$ for all $S$.
A hard-core constraint, stemming from the mapping of spins to bosons, is not included in our theory.
While it is well-known how to treat the hard-core constraint in an exact way for $S=1/2$,\cite{nikuni95} this is not the case for $S>1/2$.

\section{DMRG results for periodic boundary conditions}
\label{app:dmrg}

The DMRG results presented in the main text were computed for open boundary conditions (OBC), except for those in Fig.~\ref{fig:svn}
where we used periodic boundary conditions (PBC).
Alternatively, one may compute all the data, {\it e.g.}, the magnetization curves $M(h)$ using PBC. 
This approach is generally expected to suffer from (i)
slower convergence with respect to the number of states kept in the DMRG runs \cite{schollwoeck05}
and (ii) large finite-size effects due to the incommensurability in the problem. We here demonstrate that
all main features can be seen with both OBC and PBC, namely the existence of the
meta\-magnetic jump and  the Haldane gap.
Moreover, the quantitative results are comparable, except for the expected finite-size effects
due to the incommensurability.

\begin{figure}[tb!]
\includegraphics[width=0.9\columnwidth]{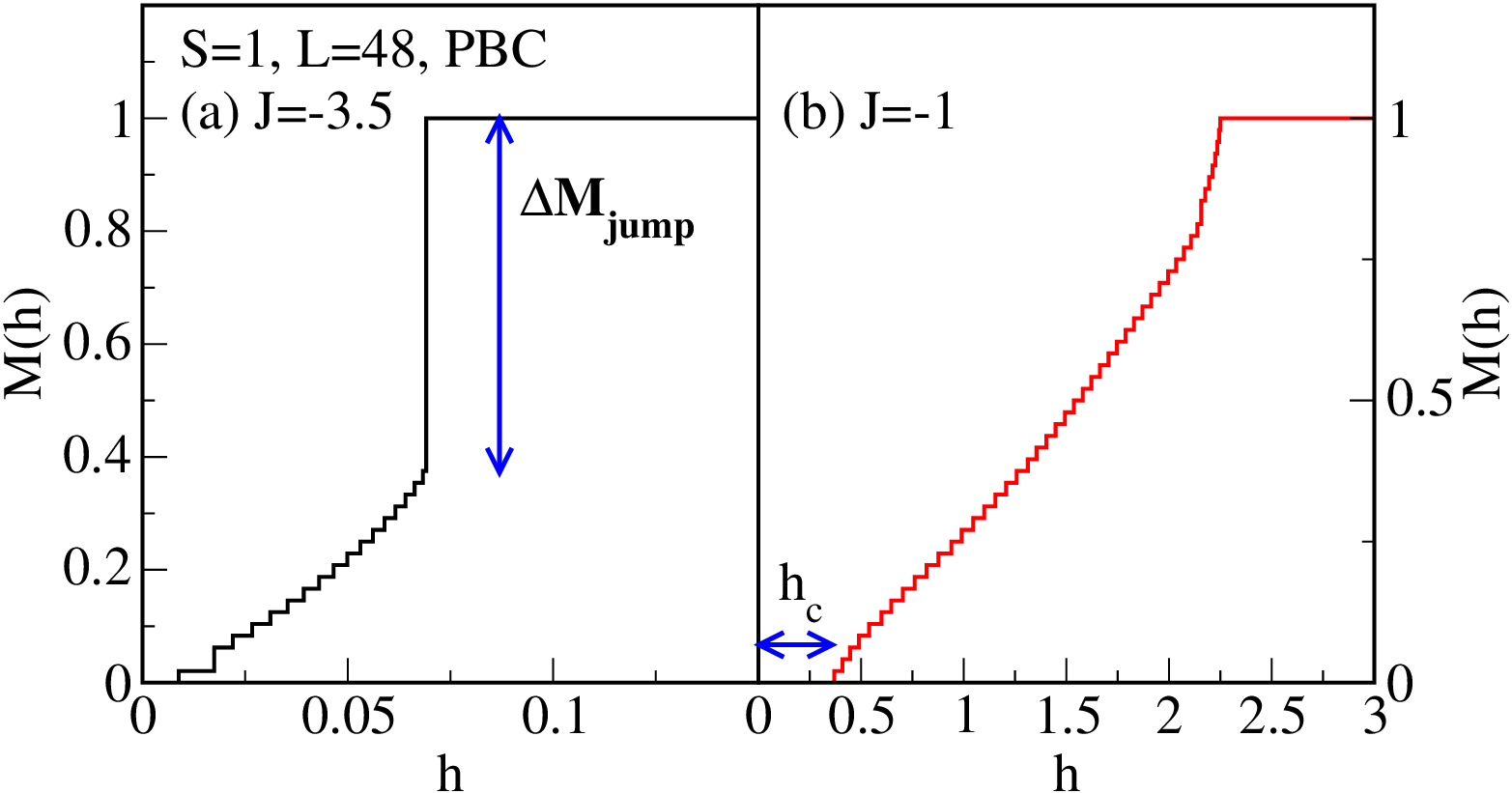}
\caption{(Color online)  Magnetization curves $M(h)$ for $S=1$ (a) $J=-3.5$  and
(b) $J=-1$ ($L=48$),
calculated with DMRG on systems with periodic boundary conditions. In panel (a), we clearly see the metamagnetic
jump of height $\Delta M_{\mathrm{jump}}$ and in panel (b), the Haldane gap that defines $h_c$ shows up as a zero-field magnetization
plateau.}
\label{fig:mag_curve_pbc}
\end{figure}

\subsection{Magnetization curves from periodic boundary conditions for $S=1$}

First, we discuss some examples of magnetization curves obtained for PBC.
We kept up to $m=1600$ states for the PBC DMRG computations presented here.
 In addition, we used exact diagonalization for those sectors which are
 sufficiently small. Therefore, in particular the data close to the saturation
 field are free of truncation errors.

Figure~\ref{fig:mag_curve_pbc} shows the magnetization curves for $J=-3.5$ [panel (a)] and $J=-1$ [panel (b)] for $L=48$ and PBC.
As in the results for OBC, we observe the presence of the metamagnetic jump (here in the case of
$J=-3.5$) and the Haldane gap (see the $J=-1$ curve), which manifests itself as a plateau in the magnetization curve at $M=0$.
The latter defines the critical field $h_c$ that separates the gapped Double-Haldane phase from the gapless finite-$M$
phase (compare Fig.~\ref{fig:svn}).

Note that at intermediate $M$, the PBC data show spurious small steps with $\Delta S^z>1$. We have checked that these features
disappear as one goes to larger system sizes.

\begin{figure}[tb!]
\includegraphics[width=0.9\columnwidth]{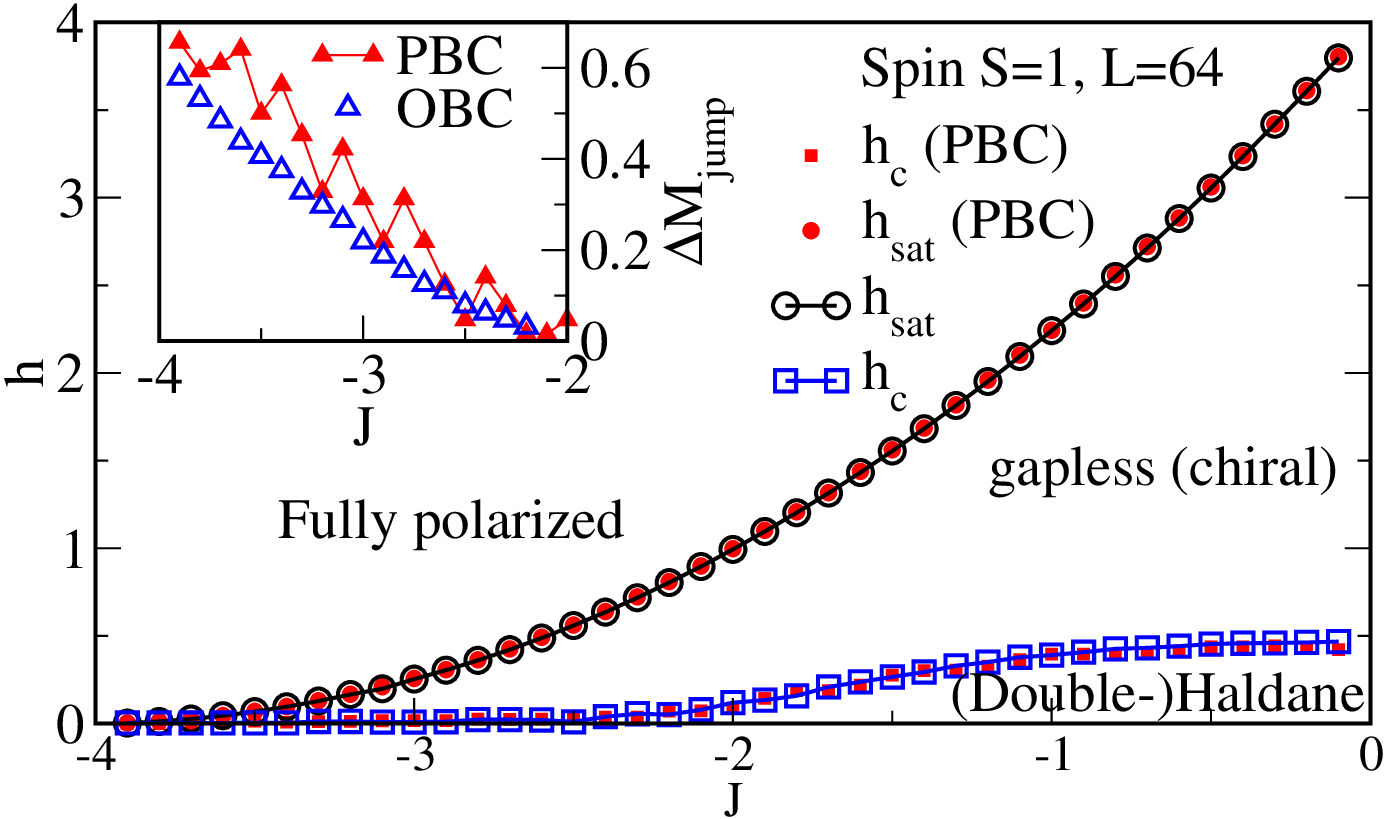}
\caption{(Color online)
Phase diagram of the frustrated ferromagnetic $S=1$ chain, comparing DMRG results from systems
with OBC (open symbols) to results from systems with PBC (solid symbols). $L=64$ in all
cases. Lines are guides to the eye.}
\label{fig:phase_diag_pbc}
\end{figure}

\subsection{Comparison of OBC vs PBC for $S=1$}

Figure~\ref{fig:phase_diag_pbc} contains the results for the Haldane gap that defines the critical field $h_c$ separating
the gapped zero-field phase from the gapless phase at finite magnetizations, the saturation field $h_{\mathrm{sat}}$, and the jump height
(inset), comparing data for OBC (open symbols) with data from PBC (solid symbols).
Here we choose a chain length of $L=64$ for both PBC and OBC.

For PBC we can determine the spin gap from
\begin{equation}
h_c^{\text{PBC}} = E(S^z=1) - E(S^z=0) \, ,
\label{eq:HcPBC}
\end{equation}
where, as in Eq.~\eqref{eq:HcOBC},
$E(S^z)$ is the ground-state energy in a sector with a given total $S^z$.
The good agreement between the OBC and PBC results for $h_c$
in Fig.~\ref{fig:phase_diag_pbc} confirms that it is indeed appropriate
to use Eq.~\eqref{eq:HcOBC} for OBC.

The PBC data for $\Delta M_{\mathrm{jump}}$ in Fig.~\ref{fig:phase_diag_pbc}
suffer from the presence of several peaks, resulting in
a non-monotonic dependence on $J$. This is due to the incommensurability in the finite-magnetization region,
which is incompatible with the lattice vectors of a system with periodic boundary conditions that is
translationally invariant. Therefore, in the main text we focus on the discussion of DMRG data from
systems with OBC. The results from OBC and PBC data for the saturation field $h_{\mathrm{sat}}$, however, agree very well with each other.

\end{document}